\documentclass[a4paper,11pt]{article}
\usepackage{pos}

\usepackage{mciteplus}
\usepackage{tikz}
\usepackage{slashed}
\usepackage{mathtools}
\newcommand{\rmd}{\mathrm{d}}

\title{Zero-jettiness beam functions at \texorpdfstring{N$^3$LO}{N3LO}}
\author*[a,b]{Arnd Behring}

\affiliation[a]{Institut für Theoretische Teilchenphysik,
  Karlsruher Institut für Technologie, \\
  76128 Karlsruhe, Germany}

\affiliation[b]{Theoretical Physics Department, CERN, \\
   1211 Geneva 23, Switzerland}

\emailAdd{arnd.behring@cern.ch}

\abstract{%
  The zero-jettiness beam functions describe collinear emissions from initial
  state legs and appear in the factorisation theorem for cross sections in the
  limit of small zero-jettiness. They are an important building block for
  slicing schemes for colour-singlet production at hadron colliders. We report
  on our ongoing calculation of this quantity at next-to-next-to-next-to-leading
  order (N$^3$LO) in QCD, highlighting in particular the aspects of partial
  fraction relations and the calculation of master integrals.
}

\FullConference{%
  Loops and Legs in Quantum Field Theory - LL2022,\\
  25-30 April, 2022\\
  Ettal, Germany
}

\begin{document}
\maketitle

\section{Introduction}
With the LHC collecting more and more data, there has been a continuous push for
higher precision, both in experimental measurements and in theoretical
predictions. Over the last few years, first N$^3$LO QCD predictions for hadron
collider observables have started to appear -- first for inclusive quantities
\cite{Anastasiou:2015vya,Anastasiou:2016cez,Duhr:2019kwi,Duhr:2020kzd,%
Chen:2019lzz,Dreyer:2018qbw,Duhr:2020sdp,Duhr:2020seh,Duhr:2021vwj},
and more recently also for differential distributions
\cite{Cieri:2018oms,Dulat:2018bfe,Chen:2021vtu,Billis:2021ecs,Chen:2021isd,%
Camarda:2021ict,Chen:2022cgv,Chen:2022lwc,Neumann:2022lft}.
A first obvious target for such high-precision calculations are colour-singlet
production processes, e.g., Higgs boson production or Drell-Yan processes. They
serve as standard candles for the Standard Model (SM), but they are also crucial
for many important measurements, including determinations of SM parameters,
such as the strong coupling constant $\alpha_s$, parton distribution
functions, the weak boson masses etc.

Differential calculations beyond leading order require a prescription for how to
deal with infrared (IR) singularities, which appear due to massless particles
becoming soft or collinear to each other. They are typically regulated in
dimensional regularisation and appear as poles in the dimensional regulator
$\epsilon$. For appropriate (IR-safe) observables, these poles cancel upon
combining real and virtual contributions. However, since they appear in real
emission contributions only after integrating over the phase space of unresolved
particles, this poses a problem for numerical calculations, which are typically
required for differential predictions. It is therefore necessary to define a
scheme to extract and cancel these singularities before the numerical
phase-space integration can be performed.
There are many different schemes to deal with IR singularities available in the
literature. Many of them fall in two broad categories: slicing schemes and
subtraction schemes. At N$^3$LO, calculations based on slicing schemes seem more
achievable at the moment due to the possibility to build on existing NNLO
calculations for processes with higher multiplicities.

The idea of phase-space slicing is to subdivide the phase space into a
fully-unresolved and a resolved region. To achieve this, one often uses an
observable which is sensitive to the fully-unresolved region and then defines a
threshold of this observable that separates the two regions. Below the
threshold lies the fully-unresolved region which contains the most complicated
singularity structure. However, if the slicing variables is chosen well, the
cross section in this region significantly simplifies thanks to, e.g.,
factorisation theorems. This approximation captures the most singular behaviour
and the integration over the unresolved phase space can then be performed
analytically, thereby exposing the poles in $\epsilon$ explicitly.
Above the threshold, the cross section is screened from the fully unresolved
configuration due to the threshold and therefore corresponds to the same process
at one order lower in perturbation theory but with an additional hard parton.
For colour-singlet production at N$^3$LO, for example, the region above the
threshold can equivalently be interpreted as colour-singlet plus one jet
production at NNLO, where the threshold of the slicing observable determines
the cuts on the jet. The IR singularities in this region can be dealt with
using available NNLO subtraction schemes. If the threshold is chosen small
enough, the sum of the two phase-space regions should become independent of
the value of the threshold parameter, in spite of the approximation made
below the threshold.

There are several possible slicing variables. The two most common ones for
colour-singlet production are the transverse momentum $q_T$ of the
colour-singlet state \cite{Catani:2007vq} and zero-jettiness $\tau$
\cite{Gaunt:2015pea}. The latter is defined as \cite{Stewart:2010tn}
\begin{align}
  \tau &= \sum_j \min_{i \in \{1,2\}} \frac{p_i \cdot k_j}{Q_j}
  \,,
\end{align}
where $j$ labels final state partons with momenta $k_j$ and the momenta $p_i$,
$i \in \{1,2\}$, are the momenta of the incoming partons. The $Q_i$ are
normalisation scales. The definition ensures that the zero-jettiness variable
$\tau$ vanishes exactly when all emission momenta $k_j$ become soft or collinear
to one of the initial state partons, i.e.~when the configuration approaches the
fully-unresolved limit.

For a zero-jettiness slicing scheme, we are interested in finding an
approximation of the cross section in the fully-unresolved region. In the limit
$\tau \to 0$, there exists a factorisation theorem
\cite{Stewart:2010tn,Stewart:2009yx,Jouttenus:2011wh}, which was proven in
soft-collinear effective theory (SCET), that the cross-section factorises
as follows,
\begin{align}
  \lim_{\tau \to 0} \sigma
    &= B \otimes B \otimes S \otimes H \otimes \sigma_{\text{LO}}
       +\mathcal{O}(\tau)
  \,.
\end{align}
Here, $H$ is the hard function, which describes corrections to the hard process.
The soft function $S$ describes the effects of soft, non-collinear gluons or
$q\bar{q}$ pairs. It is known to NNLO and the calculation of the
N$^3$LO contribution is currently underway. Finally, there are the beam
functions $B$, which describe collinear emissions off initial state partons.
First N$^3$LO results for the beam functions in singular limits were published
in Ref.~\cite{Billis:2019vxg} and in the large $N_c \sim n_f$ limit
in Refs.~\cite{Melnikov:2018jxb,Melnikov:2019pdm,Behring:2019quf}. The full
result was published in Ref.~\cite{Ebert:2020unb}. Here we report on the ongoing
independent calculation of these beam functions at N$^3$LO, which is certainly
warranted given the complexity of this problem.

\section{Beam functions for zero-jettiness}
The beam functions for zero-jettiness are non-perturbative objects which depend
on the longitudinal momentum fraction $x$ and the transverse virtuality $t =
-((p^{*})^2-k_\perp^2)$, where $p^*$ is the momentum of the parton entering the
hard process and $k_\perp$ is the transverse component of the sum of momenta of
all emitted partons.
They are related to traditional collinear PDFs $f_i(x,\mu)$ via a convolution,
\begin{align}
  B_i(t,x,\mu)
    &= \int_0^1 \rmd z \sum_{j \in \{q,\bar{q},g\}} I_{ij}(t,z,\mu)
         f_j\left(\frac{x}{z},\mu\right)
       +\mathcal{O}\left(\frac{\Lambda_{\text{QCD}}^2}{t}\right)
  \,,
\end{align}
where $i \in \{q,\bar{q},g\}$ is a flavour index, and $I_{ij}(t,z,\mu)$ are
perturbatively calculable matching coefficients. At leading order the matching
coefficient reads $I_{ij}^\text{LO}(t,z,\mu) = \delta(1-z) \delta(t)
\delta_{ij}$, which implies that the leading order beam functions correspond to
$B_i^\text{LO}(t,x,\mu) = f_i(x,\mu) \delta(t)$. The goal is to compute the
matching coefficients at N$^3$LO in QCD.

In order to understand how to calculate the matching coefficients, it is useful
to recall that both the zero-jettiness beam function and the PDFs are defined as
matrix elements between proton states, $|P(p)\rangle$, of certain operators in
SCET, which we write schematically as (the exact definition of the operators
$\mathcal{O}_i$ and $\mathcal{Q}_i$ is not important for the following
discussion)
\begin{align}
  B_i &\sim \langle P(p) | \mathcal{O}_i(t,x p,\mu) | P(p) \rangle \,, &
  f_i &\sim \langle P(p) | \mathcal{Q}_j(x' p, \mu) | P(p) \rangle
  \,.
\end{align}
The matching relation, mentioned above, arises from an operator product
expansion, which is of course valid for arbitrary external states. Thus, the
matching coefficients are independent of the external states which means that we
can choose massless partons as external states to simplify the calculation. To
this end, we define partonic beam functions $B_{ik}$ and partonic PDFs $f_{jk}$
by replacing the external states, i.e.,
\begin{align}
  B_{ik} &\sim \langle p_k(p) | \mathcal{O}_i(t,x p,\mu) | p_k(p) \rangle \,, &
  f_{jk} &\sim \langle p_k(p) | \mathcal{Q}_j(x' p, \mu) | p_k(p) \rangle
  \,.
\end{align}
The partonic beam functions and PDFs carry an additional flavour index $k$,
which corresponds to the flavour of the external parton state $|p_k(p)\rangle$.
Since the partonic PDFs are expressible in terms of the DGLAP splitting
functions, the matching coefficients can be determined by calculating the
partonic beam functions $B_{ij}$ in perturbative QCD.

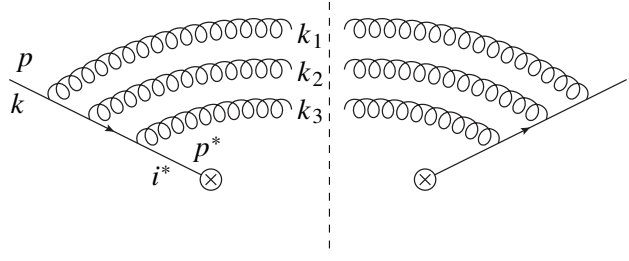
\begin{figure}
  \centering
  \begin{tikzpicture}
    % Squared line
    \draw[dashed] (4.5,0) -- (4.5,-3.3);
    % Squared amplitude
    \node[anchor=north west] at (0,0)
      {\includegraphics[scale=0.47]{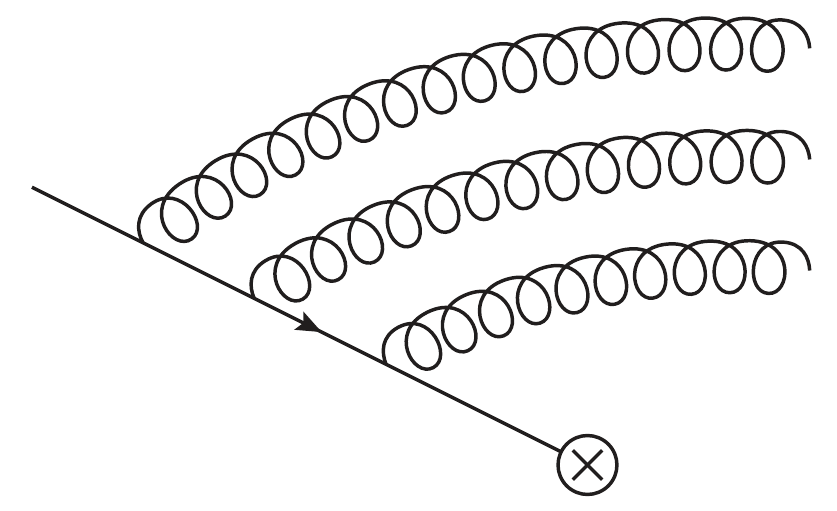}};
    \node[anchor=north west,xscale=-1] at (8.7,0)
      {\includegraphics[scale=0.47]{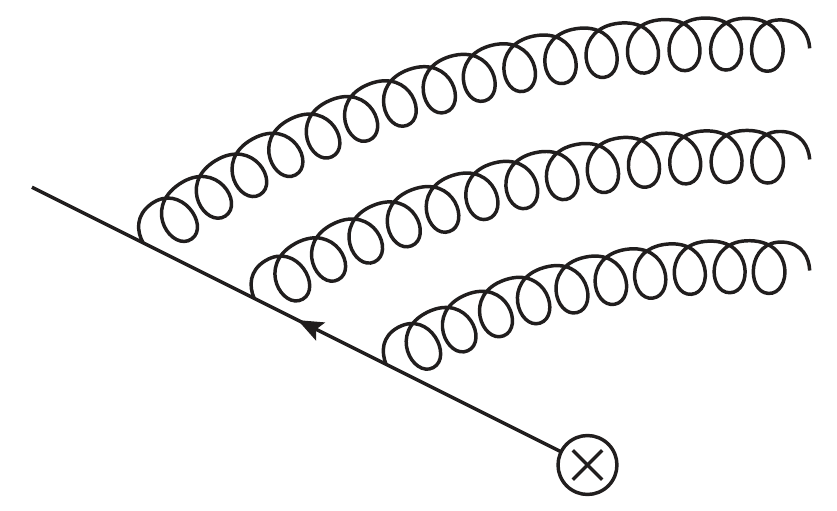}};
    % Labels
    \node at (0.50,-0.80) {$p$};
    \node at (2.90,-1.95) {$p^*$};
    \node at (4.25,-0.45) {$k_1$};
    \node at (4.25,-0.95) {$k_2$};
    \node at (4.25,-1.45) {$k_3$};
    \node at (0.40,-1.35) {$k$};
    \node at (2.30,-2.30) {$i^*$};
  \end{tikzpicture}
  \caption{Diagram from the triple real contribution of the partonic
    beam function $B_{ik}$. The external parton carries momentum $p$
    and flavour $k$ and the off-shell parton entering the hard process
    carries momentum $p^*$ and flavour $i^*$. We label the momenta of
    the emitted partons by $k_i$.}
  \label{fig:diagram-kinematics}
\end{figure}
Our calculation is based on the observation made in
Ref.~\cite{Ritzmann:2014mka}, that the partonic beam functions can be calculated
from the collinear limits of QCD amplitudes via
\begin{align}
  B_{ik}^{\text{bare}}
    &\sim \sum_{n_R} \int\prod_{i=1}^{n_R}
            \frac{\rmd^d k_i}{(2\pi)^{d-1}} \delta_+(k_i^2) \,
            \delta\left(2 p \cdot k_{1 \dots n_R} - \frac{t}{z}\right)
            \delta\left(\frac{2 \bar{p} \cdot k_{1 \dots n_R}}{s} - (1-z)\right)
            \frac{\hat{C}_p |M(p,\bar{p},\{k_i\})|^2}{|M_0(z p,\bar{p})|^2}
  \,,
  \label{eq:bare-beam}
\end{align}
where $n_R$ is the number of additional partons radiated, $k_{1 \dots n_R} = k_1
+ \dots + k_{n_R}$ is the sum of their momenta and $\bar{p} = (p^0,-\vec{p})$ is
used as the reference vector which is back-to-back with the incoming momentum
$p$. The operator $\hat{C}_p$ extracts the collinear $k_i \parallel p$ limit
from the squared matrix element $|M(p,\bar{p},\{k_i\})|^2$. Since we divide by
the corresponding reduced matrix element $|M_0(zp,\bar{p})|^2$, we essentially
obtain the splitting function from the ratio $\hat{C}_p
|M(p,\bar{p},\{k_i\})|^2/|M_0(z p,\bar{p})|^2$ and we have to integrate over the
phase space of the emission momenta, which is constrained by the two delta
functions that introduce the observables $t$ and $z$. We treat the delta
functions using reverse unitarity \cite{Anastasiou:2002yz} by expressing them as
cut propagators. To construct the splitting functions, we follow
Ref.~\cite{Catani:1999ss} and generate diagrams for an on-shell parton with
momentum $p$ and flavour $k$ going to an off-shell parton with momentum $p^*$
and flavour $i$ and additional emissions of $n_R$ partons. The off-shell parton
is then connected to a suitable projector which extracts the collinear
behaviour. For example, for quarks the Feynman rule for the operator reads
\begin{align}
  p^* \vcenter{\hbox{\includegraphics[width=6em]{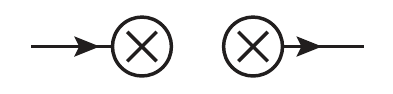}}}
    &\rightarrow \frac{\bar{p}_\mu \gamma^\mu}{4 N_c p^* \cdot \bar{p}}
  \,.
\end{align}
For the rest of the diagrams we use standard QCD Feynman rules, with the only
further peculiarity that we have to work in axial gauge, where the gluon
propagator reads
\begin{align}
  k \vcenter{\hbox{\includegraphics[width=6em]{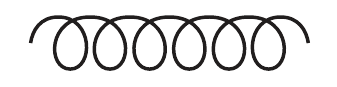}}}
    &\rightarrow \frac{i \left(-g^{\mu\nu}
                       + \frac{k^\mu \bar{p}^\nu + k^\nu \bar{p}^\mu}{%
                               k \cdot \bar{p}}
                         \right)}{k^2 + i0^+}
  \,.
  \label{eq:axial-gluon-prop}
\end{align}
This ensures that the leading collinear behaviour can be obtained by considering
only emissions off a single leg. The second term in the numerator in
Eq.~\eqref{eq:axial-gluon-prop} introduces linear propagators.

\section{Calculation}
In general, the calculation of the bare beam function follows a mostly standard
chain of calculational steps. We generate diagrams for the process $k \to i^* +
\text{emissions}$, interfere them with the appropriate, complex conjugated
diagrams and perform the Dirac and colour algebra. We then apply partial
fractioning to be able to map the scalar integrals to integral families, which
we further reduce to master integrals using integration-by-parts identities.
Afterwards, we use partial fractioning once more on the master integrals, in
order to find relations between master integrals from different integral
families and thereby reduce the overall number of master integrals. To solve the
master integrals we use the method of differential equations and we fix the
integration constants in the limit $z \to 1$. Finally, we apply renormalisation
and IR subtractions in order to translate the bare partonic beam functions to
partonic beam functions, from which we can finally extract the matching
coefficients.

\begin{figure}
  \centering
  \includegraphics[width=0.32\textwidth]{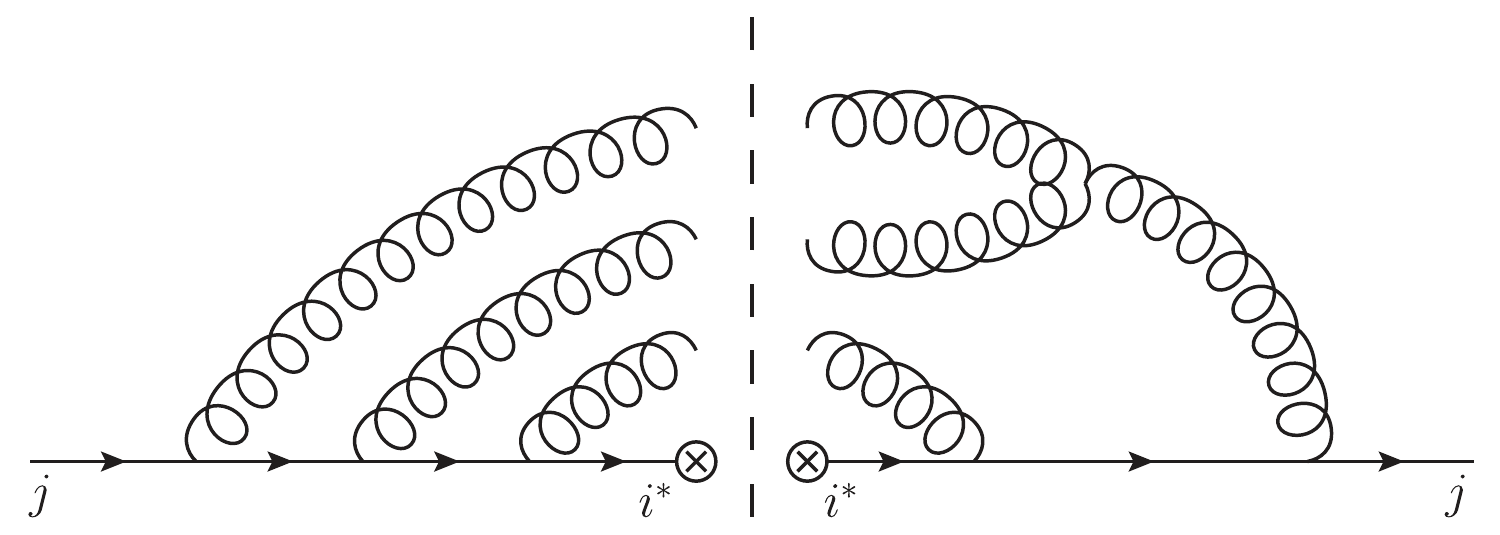}
  \includegraphics[width=0.32\textwidth]{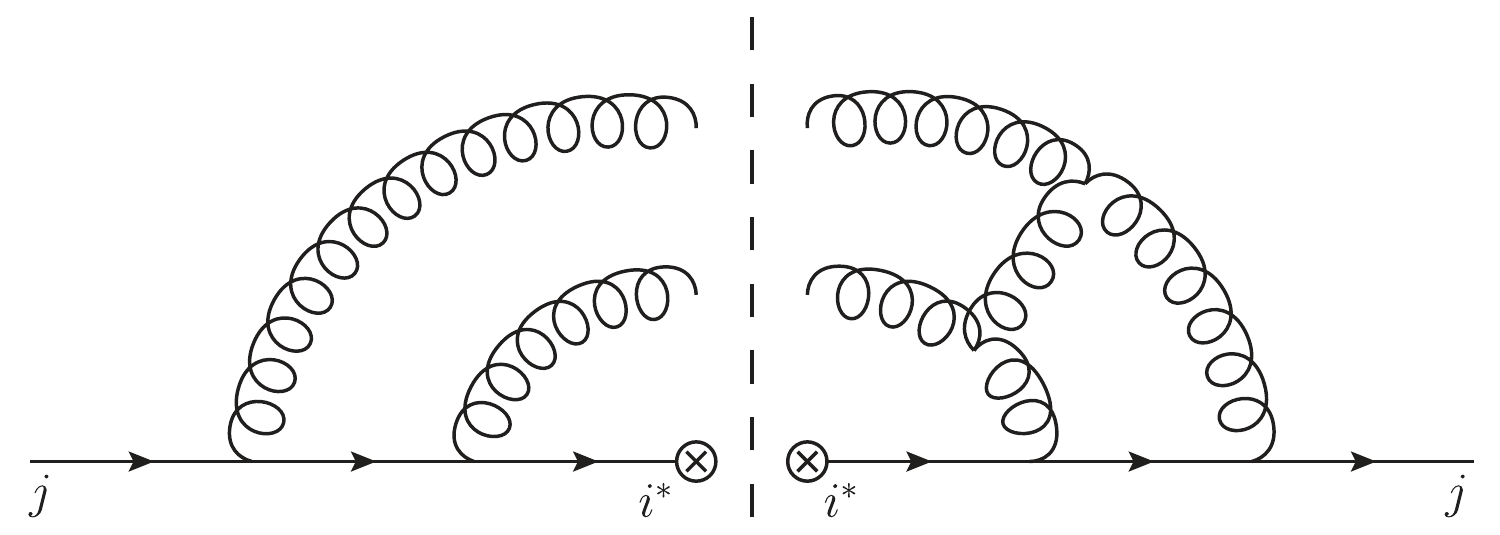}
  \includegraphics[width=0.32\textwidth]{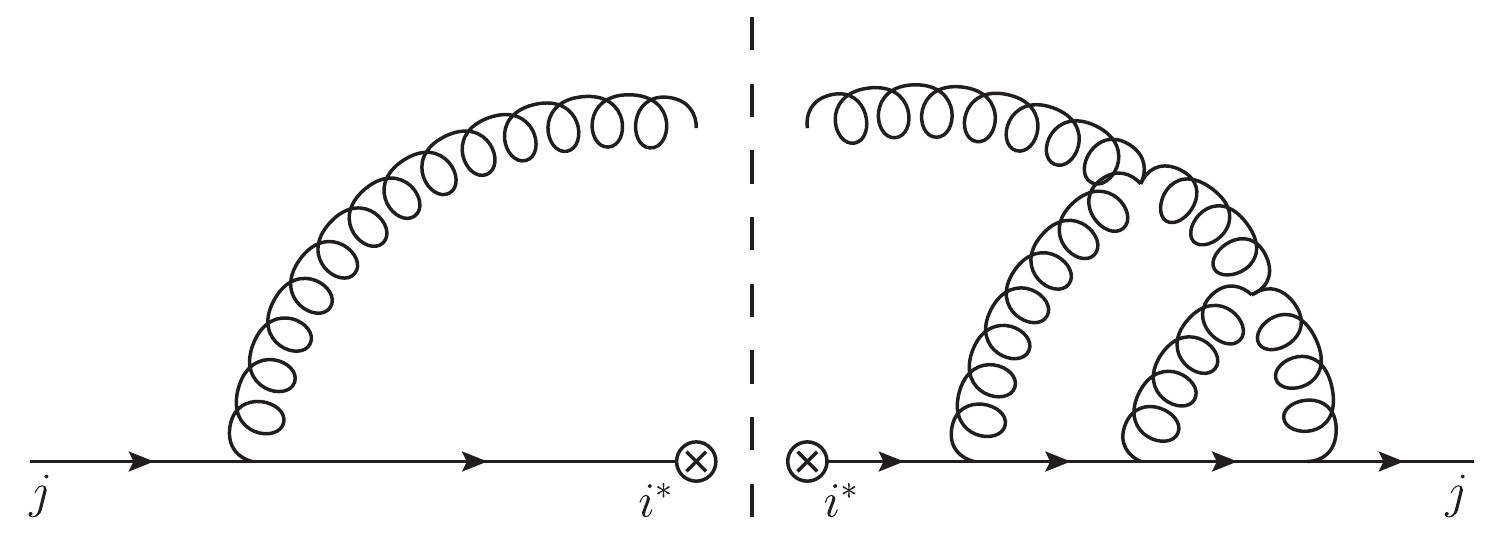}
  \caption{Example diagrams for the triple-real (RRR),
           double-real-single-virtual (RRV) and
           single-real-double-virtual (RVV) contributions.}
  \label{fig:example-diags}
\end{figure}
We organise the calculation according to the number of virtual loop integrations
that the different contributions require. There are triple-real (RRR),
double-real-single-virtual (RRV) and single-real-double-virtual (RVV)
contributions, but there are no contributions from triple-virtual diagrams to
the beam functions. Examples for diagrams appearing in these three contributions
are shown in Fig.~\ref{fig:example-diags}. The calculation of the RRR and RRV
contributions follows the steps described above. This amounts to having to
calculate about 450 three-loop master integrals. The solutions to these master
integrals depend on iterated integrals over an alphabet containing both linear
letters,
\begin{align}
  f_a(z) &= \frac{1}{z-a}
  \,, &
  \text{where~}
  a &\in \left\{
    0, \pm 1, \pm 2, \pm 4, \pm 2i, \exp\left(\pm i \pi \tfrac{2}{3}\right)
  \right\}
  \,,
\end{align}
and square-root valued letters
\begin{align}
  \left\{
    \frac{1}{\sqrt{z (4-z)}},
    \frac{1}{\sqrt{z (4+z)}},
    \frac{1}{\sqrt{4 + z^2}},
    \frac{1}{z \sqrt{4 + z^2}}
  \right\}
  \,.
\end{align}
The RVV contribution, on the other hand, we calculate using the results for the
two-loop splitting functions published in the literature
\cite{Duhr:2013msa,Duhr:2014nda}.

In the following, we would like to highlight two particular aspects of the
calculation: the use of partial fractioning and the calculation of the master
integrals. We have to employ partial fractioning due to linear relations
between propagators, which are, for example, introduced by the delta functions
that describe the observables. An example from the RRV contribution for such a
linear relation is
\begin{align}
  2 (k_1 + k_2) \cdot \bar{p} &= s (1 - z)
  \,,
\end{align}
which gives rise to partial fractioning identities such as
\begin{align}
  \frac{1}{(k_1 \cdot \bar{p}) (k_2 \cdot \bar{p})}
    &= \frac{2}{s (1-z)} \left[\frac{1}{k_1 \cdot \bar{p}}
         +\frac{1}{k_2 \cdot \bar{p}}\right]
  \,.
\end{align}
In order to map the scalar integrals to integral families, all propagators
appearing in the integrals have to be linearly independent. Therefore, we have
to apply partial fractioning to map integrals with linearly dependent
propagators to sums of integrals with linearly independent propagators. Similar
to linear relations induced by the delta functions, each gluon propagator in
the axial gauge introduces a linear propagator $(k \cdot \bar{p})^{-1}$ which
also leads to linearly dependent propagators: Since there are three linearly
independent scalar products involving $\bar{p}$ and since every integral has to
contain at least the cut propagator corresponding to $\delta(2 k_{1\dots n_R}
\cdot \bar{p} - s(1-z))$ it is easy so see that many of the integrals arising
from diagrams with more than two gluons require partial fractioning in order to
be mapped to integral families. To systematically apply the partial fractioning
identities, we use Gröbner bases, as described in
Refs.~\cite{Pak:2011xt,Hoff:2015kub} (see also
Refs.~\cite{Meyer:2017joq,Abreu:2019odu,Heller:2021qkz} for related work).

In addition, we also apply partial fractioning to derive relations between
master integrals from different integral families. An example for such a
relation from the RRR contribution is
\begin{align}
  \MoveEqLeft{\overbrace{\int \frac{\rmd \Phi_B}{(k_1-p)^2 (k_{13}-p)^2 (k_3 \cdot \bar{p}) (k_{13} \cdot \bar{p})}}^{=I^\text{T29}_{3391}}}
  \notag \\
    &=  \underbrace{\int \frac{\rmd \Phi_B}{(k_1-p)^2 (k_{12}-p)^2 (k_1 \cdot \bar{p}) (k_2 \cdot \bar{p})}}_{=I^\text{T1}_{3263}}
       -\underbrace{\int \frac{\rmd \Phi_B}{(k_1-p)^2 (k_{13}-p)^2 (k_1 \cdot \bar{p}) (k_{13} \cdot \bar{p})}}_{=I^\text{T28}_{3291}}
  \,,
  \label{eq:pfrel}
\end{align}
where $\rmd \Phi_B$ is the phase-space measure shown in Eq.~\eqref{eq:bare-beam}
and $I_s^f$ represents an integral from integral family $f$ and sector $s$.
The relation arises from the linear dependence of the linear propagators in the
three integrals, which is immediately clear at the integrand level after
relabelling $k_2 \leftrightarrow k_3$. Since each integral family can only
contain linearly independent propagators, the three master integrals must belong
to different integral families. Applying these relations significantly reduces
the number of master integrals that have to be calculated. We construct these
relations by first generating a list of seed integrals from all sectors of all
integral families, computing a Gröbner basis for the set of all appearing
propagators and applying it to all seed integrals. Subsequently, we apply IBP
relations to reduce all appearing integrals to master integrals and solve the
resulting linear system. This approach is rather brute force and leaves open
the question whether all possible relations are found in this way. The relation
in Eq.~\eqref{eq:pfrel} already requires the shift $k_2 \leftrightarrow k_3$,
which shows that in each term of the partial fractioning relation, there exists
a freedom to use symmetries of the Feynman integrals such as, for example, loop
momentum shifts. It stands to reason that there might be partial fractioning
relations which cannot be found by the procedure described above since the
construction of the Gröbner basis is based on one fixed choice for the momentum
space representation of the propagators. It would certainly be interesting to
investigate in the future if this construction can be further refined.

The second aspect we would like to discuss in some more detail concerns the
calculation of the master integrals. In principle, the integrals depend on
three kinematic variables: the centre of mass energy squared $s$, the transverse
virtuality $t$ and the longitudinal momentum fraction $z$. By rescaling $k_i =
\tilde{k}_i \sqrt{t}$, $p = \tilde{p} \sqrt{t}$ and $\bar{p} = \tilde{\bar{p}} s
/ \sqrt{t}$, the dependence on $s$ and $t$ can be scaled out, e.g.,
\begin{align}
  I^\text{RRV1}_{n_1,\dots,n_8}(s,t,z)
    &= s^{-1-n_7-n_8} t^{3-3\epsilon-(n_1+\dots+n_6)} I^\text{RRV1}_{n_1,\dots,n_8}(z)
  \,,
\end{align}
and the only remaining non-trivial dependence is that on $z$. Thus, we derive
differential equations in $z$ for the master integrals and fix the boundary
conditions by computing the integrals in the limit $z \to 1$. To calculate the
integrals in this limit, we use a number of different techniques, including
deriving constraints between different integration constants from the analytic
structure of the integrals, direct calculation, auxiliary differential equations
and mapping the limit of RRR integrals to threshold integrals that appear in Higgs
production ($gg \to H$).

Let us briefly sketch the main ideas behind the last approach. To calculate the
soft region of beam function master integral $I(s,t,z)$ in the limit $z \to 1$ we
first introduce a new integral $\mathcal{B}(s,t,z)$, in which the propagators
appearing in $I(s,t,z)$ are replaced by their eikonal approximations (but the
integration measure is the same). In the limit $z \to 1$ the two integrals agree,
$\lim_{z \to 1} I(s,t,z) = \lim_{z \to 1} \mathcal{B}(s,t,z)$. Since the integrand
of $\mathcal{B}(s,t,z)$ is homogeneous in the integration momenta, also the
dependence on $z$ can be scaled out,
\begin{align}
  \mathcal{B}(s,t,z)
    &= s^{e_s} t^{e_t} z^{-e_t} (1-z)^{e_{\bar{z}}} \tilde{\mathcal{B}}
  \,,
  \label{eq:soft-scaling}
\end{align}
where $e_s$, $e_t$ and $e_{\bar{z}}$ are exponents that can be read off from the
integrand and integration measure and $\tilde{\mathcal{B}}$ is a numerical
constant. Next, we compare the phase-space measure for the beam function to the
phase-space measure for Higgs production in the threshold limit. The important
aspect are the delta functions that appear. For the beam function they read
\begin{align}
  \mathcal{B}(s,t,z)
    &\sim \delta\left(2 k_{123} \cdot p - \frac{t}{z}\right)
          \delta\left(2 k_{123} \cdot \bar{p} - s (1-z)\right)
  \,,
\end{align}
while the phase-space measure for Higgs production in the threshold limit
contains
\begin{align}
  \mathcal{H}(s)
    &\sim \delta\left(2 k_{123} \cdot (p + \bar{p}) - s\right)
  \,.
\end{align}
The central idea is now to choose a special kinematic point, $t = s z^2$, and to
integrate over $z \in [0,1]$,
\begin{align}
  \int_0^1 \rmd z \, \mathcal{B}(s, s z^2, z)
    &= \int_0^1 \rmd z \, \delta(2 k_{123} \cdot p - s z)
         \delta(2 k_{123} \cdot \bar{p} - s(1-z))
     = \frac{1}{s} \mathcal{H}(s)
  \,.
\end{align}
Combining this with the information from Eq.~\eqref{eq:soft-scaling}, we find
\begin{align}
  \frac{1}{s} \mathcal{H}(s)
    &= s^{e_s+e_t} B(e_t+1,e_{\bar{z}}+1) \tilde{\mathcal{B}}
  \,,
\end{align}
where $B(a,b) = \Gamma(a)\Gamma(b)/\Gamma(a+b)$ is the Euler Beta function.
This shows that the numerical constant $\tilde{\mathcal{B}}$ which determines
the limit $z \to 1$ of the beam function integrals is expressible in terms of
Higgs production threshold integrals $\mathcal{H}(s)$. This connection is
helpful because it allows to further relate the boundary constants by using IBP
relations for the Higgs threshold integrals and because there are results
available for these integrals in the literature \cite{Anastasiou:2013srw,%
Anastasiou:2015yha,Duhr:2022cob}. We discovered that the integrals which we
calculated for the RRR contribution \cite{Melnikov:2019pdm} covers 9 of the 10
master integrals computed for Ref.~\cite{Anastasiou:2013srw}, including the
most complicated one from that reference, $\mathcal{F}_9$. Similarly, for the
RRV contribution, our boundary constants cover several of the soft region
master integrals from Ref.~\cite{Anastasiou:2015yha}, including the most
complicated one, $\mathcal{M}_{13}$. We note that this implies that it should
be possible to use our results to independently redo the calculation of the
N$^3$LO QCD corrections to the process $gg \to H$ in the threshold region.

\section{Results}
Overall there are five independent matching coefficients: $I_{q_i q_j}$, $I_{q
g}$, $I_{g q}$, $I_{gg}$ and $I_{q_i \bar{q}_j}$. Matching coefficients for
other possible flavour combinations can be derived from the given list using
crossing symmetries.
We have published first N$^3$LO results for $I_{q_i q_j}$ in the large $N_c \sim
n_f$ limit in Ref.~\cite{Behring:2019quf}. The full result for all five
independent matching coefficients was published in Ref.~\cite{Ebert:2020unb} by
Ebert, Mistlberger and Vita. Based on the calculation described in this
article, we have obtained results for the matching coefficients $I_{q_i
q_j}$, $I_{q g}$ and $I_{g q}$ while the calculation of $I_{g g}$ and $I_{q_i
\bar{q}_j}$ is still in progress. For the available results, the matching
coefficients agree with the results published in Ref.~\cite{Ebert:2020unb}.
As a by-product, the pole cancellation for the matching coefficients also
cross-checks the three-loop DGLAP splitting functions \cite{Moch:2004pa,%
Vogt:2004mw,Ablinger:2014lka,Ablinger:2014vwa,Ablinger:2014nga,%
Ablinger:2017tan}. We also observe a simplification of the alphabet of the
iterated integrals which appear in the final results for the matching
coefficients compared to the alphabet observed in the master integrals. In
particular, only one of the square roots remains. The remaining letters read
\begin{align}
  \left\{
    \frac{1}{z},
    \frac{1}{z-1},
    \frac{1}{z+1},
    \frac{1}{z-2},
    \frac{1}{\sqrt{z (4-z)}}
  \right\}
  \,.
\end{align}
The full results of this calculation will be presented in a forthcoming publication.

\section{Conclusions}
We calculate the beam functions for zero-jettiness at N$^3$LO via phase-space
integrals over splitting functions in QCD. In this ongoing calculation, we have
completed the matching coefficients $I_{q_i q_j}$, $I_{q g}$ and $I_{g q}$,
confirming the results published in Ref.~\cite{Ebert:2020unb}. The calculation
of the final matching coefficients $I_{q_i \bar{q}_j}$ and $I_{gg}$ is currently
underway. In parallel, there is an effort to calculate also the soft function
for zero-jettiness at N$^3$LO. First results have recently been published in
Refs.~\cite{Baranowski:2021gxe,Baranowski:2022khd,Chen:2022cvz}. It will be
interesting to put the beam functions to use in slicing calculations for
colour-singlet production, as well as in resummation applications in the future.

\acknowledgments
The author would like to thank D.~Baranowski, K.~Melnikov, R.~Rietkerk,
L.~Tancredi and C.~Wever for the collaboration on the results discussed here and
K.~Melnikov for comments on the manuscript.
This research is partially supported by the Deutsche Forschungsgemeinschaft
(DFG, German Research Foundation) under grant 396021762 - TRR 257.

\bibliographystyle{JHEPM}
\bibliography{lit}

\end{document}